\documentclass[review=false,anonymous=false]{acmart} 

\AtBeginDocument{%
  \providecommand\BibTeX{{%
    \normalfont B\kern-0.5em{\scshape i\kern-0.25em b}\kern-0.8em\TeX}}}

\setcopyright{acmcopyright}
\copyrightyear{2018}
\acmYear{2018}
\acmDOI{10.1145/1122445.xxxxxx}

\usepackage{booktabs} 
\usepackage{url}

\usepackage{caption}
\usepackage[all]{nowidow}
\usepackage{wrapfig}
\usepackage{array}
\usepackage{arydshln}
\usepackage{tabularx}
\usepackage{multirow}
\usepackage{arydshln}
\usepackage{makecell}

\usepackage{tabularx}
\usepackage{booktabs}
\usepackage{array}
\usepackage{ragged2e}
\newcolumntype{L}[1]{>{\RaggedRight\arraybackslash}p{#1}}
\newcolumntype{Y}{>{\RaggedRight\arraybackslash}X}

\usepackage{wrapfig,lipsum,booktabs} 

\def\authnotes{1}
\newcounter{notectr}[section]
\newcommand{\thenote}{\thesubsection.\arabic{notectr}\refstepcounter{notectr}}

\newcommand{\note}[2]{$\ll$#1~\thenote: #2$\gg$}
\newcommand{\cnote}[1]{\ifnum\authnotes=1 \textcolor{blue}{\note{Comment:}{#1}}\fi}

\copyrightyear{2026}
\acmYear{2026}
\setcopyright{acmlicensed}\acmConference[CHI '26]{CHI}{Nov 5-9, 2026}{USA}
\acmBooktitle{CHI'24, Nov, 2026, USA}
\acmPrice{15.00}
\acmDOI{10.1145/3491102.43}
\acmISBN{978-1-4503-9157-3/22/04}

\begin{document}



\title[Data Patching]{Data Patching}

\author{ATM Mizanur Rahman}
\affiliation{
  \department{Computer Science}
  \institution{University of Illinois Urbana-Champaign}
  \city{Champaign}
  \state{Illinois}
  \country{USA}}
\email{amr12@illinois.edu}

\author{Syed Ishtiaque Ahmed}
\affiliation{
  \department{Computer Science}
  \institution{University of Toronto}
  \city{Toronto}
  \state{Ontario}
  \country{Canada}}
\email{ishtiaque@cs.toronto.edu}

\author{Sharifa Sultana}
\affiliation{
  \department{Computer Science}
  \institution{University of Illinois Urbana-Champaign}
  \city{Champaign}
  \state{Illinois}
  \country{USA}}
\email{sharifas@illinois.edu}

\renewcommand{\shortauthors}{Rahman et al.}

\begin{abstract}

To provide data-driven quality services to their citizens, every institution determines the acceptability of the citizen data and datafied mechanisms through institutional protocols, including preset standards and policies. However, data meeting standards at one institution might fail to meet a different institution's standards in subsequent phases of the intended citizen services due to mismatched protocols, leading to \textit{data devaluation} in the citizen service ecosystem. Terming this phenomenon \textit{cross-institutional data devaluation}, we investigate its causes and workarounds through interviews with 41 Bangladeshi participants. We found that traditional data auditing mechanisms cannot solely address data devaluation; hence, we draw on our findings and theory of cross-institutional AI audits to propose the \textit{Citizen-centered Cross-institutional Data Audit (CCDA)}. We also discuss design and policy implications of CCDA in HCI and datafied citizen services.


\end{abstract}





\keywords{Cross-institutional data devaluation, data patching, datafied citizen services, institutional seams, citizen-centered auditing}


\settopmatter{printfolios=true}

\maketitle

\section{Introduction}

In an increasingly datafied society, access to citizen services such as education, employment, healthcare, banking, and public services depends on documents and records held by institutions. Multi-institutional involvement is often necessary for carrying out everyday activities because no single institution holds everything a citizen needs to prove. As records move across institutional boundaries, they can fall through the cracks between institutions. The institution providing the record may have followed its own institutional protocol when producing it, and the institution receiving the record may also follow its own protocol when deciding whether to accept it. Both institutions may therefore function according to their own rules while the coupling between them still fails.   

We define this process as \textit{cross-institutional data devaluation}, the sociotechnical and institutional process through which information that is recognized and useful in one setting loses recognition, credibility, or practical value in another. Here, data includes documents, certificates, institutional records, professional histories, and other forms of evidence that institutions use to classify citizens and make data-driven decisions about them while delivering services. We use the term for seams between two standalone institutions, and within the same institution operating at different points in time. To understand how people respond to these failures, we draw on Rajesh Veeraraghavan's concept of patching, developed through his study of governance in India \cite{veeraraghavan2022patching}. We extend this idea to datafied citizen services and term the extra work citizens do to make data work across failed institutional connections \textit{data patching}.  


Prior work on documents, classification, and institutional authority shows that documents gain authority from institutional procedures and classifications \cite{Hull2012,BowkerStar1999,Scott1998}. Studies of Aadhaar and cross-border data migration show how state agencies and credential-evaluation bodies help records move across institutional systems \cite{SinghJackson2017,SinghJackson2021,RohanifarEtAl2026}. Infrastructure research further shows that such movement depends on shared standards, categories, gateways, and human labor \cite{StarRuhleder1996,JacksonChongthammakun2011,StarStrauss1999}. However, less is known about citizen services that involve multiple standalone institutions when no institution has control or jurisdiction over the crossing.

To investigate these institutional crossings, we conducted 41 semi-structured interviews in Bangladesh with people who had experienced difficulties making their records actionable or functional across institutional boundaries. AI and algorithm audits have become important methods for examining bias, error, discrimination, and other harmful system behavior \cite{SandvigEtAl2014,MetaxaEtAl2021AuditingAlgorithms,RajiBuolamwini2019}. Such audits work most clearly when auditors can identify a system to examine, observe a relationship between its inputs and outputs, compare its behavior against a standard, and address an actor with the authority to respond. We are motivated by the limits of this individualistic centered view of auditing for understanding harms in datafied citizen services that emerge at institutional seams. Cross-institutional data devaluation challenges these assumptions because harms may emerge from interactions among multiple technical and institutional actors rather than from a single model, database, institution, or decision \cite{selbst2019fairness, cooper2022accountability}. Auditing each institution separately may therefore show that both follow their own procedures while missing the failure in the relationship between them. We therefore draw on theory of Cross-Institutional AI Auditing \cite{SultanaAhmed2026CrossInstitutional} and extend its cross-institutional audit perspective to datafied citizen services. Our research questions are:  

\begin{itemize}

    \item[] \textit{\textbf{RQ1:} How do valid records lose recognition and value as they move across institutional boundaries in datafied citizen service ecosystems, and what structural conditions lead to data devaluation?}

    \item[] \textit{\textbf{RQ2:} What workarounds do people use to avoid data devaluation in their intended citizen services involving multi-institutions with independent protocols? How much they cohere and differ with Veeraraghavan’s patching?}

   \item[] \textit{\textbf{RQ3:} How do individualistic institutional audits fail to ensure the quality of data valuation involving patching, and what adaptations do audit mechanisms need in theory and application to address data devaluation in citizen service ecosystems?}
   
\end{itemize}

We found that independent datafied citizen services can work on their own but fail when they need to connect. Some institutions had no established connection between them, so valid records could not move, match, or be treated as equivalent across institutions. In other cases, institutions were connected, but one followed more rigid or unbending rules and gave less value to records from the other institution. We also found that some institutional connections had worked before but stopped working after political, infrastructural, or rule changes. When this data devaluation happened, citizens used workarounds such as paying gatekeepers, reproducing evidence, and assembling fragmented records. However, these workarounds sometimes failed because institutions lacked a shared standard, no one was responsible for the crossing, or the route for updating the record was closed.






We make four contributions to HCI and algorithmic accountability scholarship. First, we categorize the forms of data devaluation in citizen services that involve multiple institutions and map their causes. Second, drawing on people’s workarounds, we identify data patching that emerges when institutional connections fail. Third, we identify the limitations of existing data-audit methodology in addressing such data patching. Fourth, drawing on our findings and Cross-Institutional AI Auditing \cite{SultanaAhmed2026CrossInstitutional}, we develop the Citizen-centered Cross-institutional Data Audit (CCDA) and discuss its design and policy implications for HCI and datafied citizen services.

\section{Related Work}

\subsection{Documents, Classification, and Institutional Legitimacy}

Documents help institutions organize knowledge, rules, responsibilities, and decisions \cite{Harper1998,Hull2012,SharmaGupta2006}. Their authority comes from the institution that produced them, the procedures used to create them, and features such as signatures, stamps, seals, file formats, and official delivery channels \cite{Hull2012,SharmaGupta2006,Porter1995}. Institutions also use categories to organize people, records, and events before making decisions \cite{Scott1998,BowkerStar1999}. However, fixed categories cannot capture every part of a person's experience. If someone's history does not fit an available category, the institution may not recognize that experience \cite{StarBowker2007,BowkerStar1999}. Recognition is therefore an institutional judgment about what a record counts as \cite{MartinLynch2009}. A person can be recorded in a system but still not be recognized as eligible for a right, service, or opportunity \cite{BreckenridgeSzreter2012,CaplanTorpey2001,Sriraman2011}. HCI research on Aadhaar shows this gap. A person may be registered but still denied welfare when the system cannot match their data, categorize them correctly, or verify their identity during service delivery \cite{SinghJackson2017,SinghJackson2021}. Singh and Jackson describe this through the concept of resolution. A person may be clearly represented in one system but become ``low-resolution'' in another institutional process \cite{SinghJackson2021}. The form of a document can also affect recognition. Aadhaar authorities treated different forms of the same record as valid, but frontline officials still treated mailed letters, plastic cards, and ordinary printouts differently \cite{SinghJackson2017}.

Institutions also create common measures for comparing people, records, and experiences \cite{EspelandStevens1998}. They decide what counts as credible, legitimate, or worthy \cite{Lamont2012}. These decisions can place people and their records into unequal categories and shape access to citizen services \cite{FourcadeHealy2013,Fourcade2021}. Recent HCI research on \textit{data migration} shows a similar problem across national borders. Immigrants' educational, financial, medical, and professional histories may be reclassified, reduced, or dismissed in a new country, even when the original information remains accurate \cite{RohanifarEtAl2026}. Together, this work shows that a record can remain accurate and officially valid while losing recognition or practical value in another institutional setting. Prior work has examined this within connected systems such as Aadhaar and welfare delivery and across national borders \cite{SinghJackson2017,SinghJackson2021,RohanifarEtAl2026}. Less is known about what happens when citizen services involve multiple standalone institutions with their own standards and procedures, or when an institutional relationship changes over time. This gap motivates our first research question: How do valid records lose recognition and value as they move across institutional boundaries in datafied citizen service ecosystems, and what structural conditions lead to data devaluation?

\subsection{Institutional Seams, Infrastructure, and Breakdowns}

Infrastructure is a network of connections among technologies, standards, rules, laws, organizations, and the people who use and maintain them \cite{StarRuhleder1996,Star1999}. Infrastructure also develops over time as new technologies, rules, and practices are added to older systems \cite{StarRuhleder1996,HansethLundberg2001,PipekWulf2009,LamplandStar2009}. Connections between institutions similarly depend on shared documents and standards. Shared documents can help institutions understand the same information across different settings \cite{StarGriesemer1989}, while common standards shape how information is recorded and exchanged \cite{TimmermansEpstein2010,JacksonChongthammakun2011}. Institutions can follow their own protocols and still fail when they need to connect. Research on Thai digital government shows that agencies developed systems for their own tasks using different data structures, vendors, and document processes \cite{JacksonChongthammakun2011}. These systems could work on their own but create problems when information had to move between them. Vertesi describes such places where systems meet but do not fully connect as \textit{seamful spaces} \cite{Vertesi2014}. 

People often have to do additional work when systems do not fully connect \cite{rohanifar2021money}. \textit{Articulation work} describes the labor required to coordinate separate activities and keep a process moving \cite{SchmidtSimone1996}. This work often remains invisible in formal accounts of organizations \cite{StarStrauss1999}. In Bangladesh, infrastructural transitions can leave some people and practices outside official accounts of technological progress \cite{AhmedEtAl2015}, while local workarounds can become regular parts of technology use \cite{RangaswamySambasivan2011}. Aadhaar and digital welfare systems also depend on intermediaries, family members, brokers, and local helpers \cite{Chaudhuri2019,CarswellDeNeve2021}. Immigrants similarly rely on family and community support to make their records actionable \cite{RohanifarEtAl2026}.

Repair scholarship shows that people may create local solutions when systems fail, but they usually cannot change official standards, records, or institutional arrangements \cite{Jackson2014}. Veeraraghavan's concept of \textit{governance by patching} describes small, focused corrections to implementation failures rather than complete redesign \cite{veeraraghavan2022patching}. However, this form of patching depends on an actor with jurisdiction who can change rules, procedures, technologies, or institutional arrangements.

Prior work therefore shows how people and intermediaries perform additional work when systems break, and how actors with authority can patch institutional failures. However, we know less about what happens when the person directly affected by a cross-institutional failure has to make the connection work themselves. Their workaround may move their own case forward while leaving the underlying institutional relationship unchanged. This motivates our second research question: What workarounds do people use to avoid data devaluation in their intended citizen services involving multi-institutions with independent protocols? How much they cohere and differ with Veeraraghavan’s patching?

\subsection{Auditing and Institutional Accountability}

Audits examine whether a system or decision process follows particular claims, standards, or expectations. In HCI and algorithmic accountability, much of this work focuses on AI and computational systems \cite{MarquandHouseCollective2026,MetaxaEtAl2021AuditingAlgorithms}. Black-box audits examine patterns in a system's inputs, outputs, and behavior when auditors cannot access the source code \cite{SandvigEtAl2014,MetaxaEtAl2021AuditingAlgorithms}. This approach grew from civil-rights audit studies in areas such as housing and employment \cite{PagerWestern2012,Gaddis2018,BertrandMullainathan2004}. Algorithm audits adapted this logic to computational systems using scripts, simulated accounts, queries, and controlled datasets \cite{SandvigEtAl2014,AsplundEtAl2020}. These methods have been used to study discrimination in advertising \cite{Sweeney2013,AngwinParris2016,AliEtAl2019}, price steering \cite{HannakEtAl2014}, facial-analysis accuracy \cite{BuolamwiniGebru2018,RajiBuolamwini2019}, stereotypes in search and language-model outputs \cite{KayEtAl2015,MetaxaEtAl2021ImageSociety,ArmstrongEtAl2024}, and user-led investigations of algorithmic harms \cite{ShenEtAl2021,LamEtAl2022}.

Across this literature, several conditions make individualistic institutional audits workable. Auditors need an \textit{object} to examine, such as a model, platform, product, or decision process \cite{MetaxaEtAl2021AuditingAlgorithms}. They also need \textit{observability}, so that some relationship between inputs and outputs can be examined \cite{SandvigEtAl2014,MetaxaEtAl2021AuditingAlgorithms,Sweeney2013,BuolamwiniGebru2018,ArmstrongEtAl2024}. Audits also need a \textit{standard} for evaluating the system. The choice of standard matters because a formal audit can still miss problems such as accuracy, disability, and applicant experience \cite{GrovesEtAl2024,WrightEtAl2024}. Accountability also requires an \textit{accountable actor} who can explain or respond to a problem, such as a developer, vendor, organization, regulator, or public agency \cite{WilsonEtAl2021,YoungEtAl2022}. Auditing becomes harder when systems change over time because findings may no longer apply after a system is updated or reconfigured \cite{AngwinTobinVarner2017,AliEtAl2019}. These conditions are easier to apply when an audit focuses on one bounded system or decision process. In datafied citizen services that involve multiple institutions, however, problems may emerge at the connection between institutions. Each institution may follow its own standards and procedures while the crossing between them still fails. 

Recent work on Cross-Institutional AI Auditing addresses a similar problem by shifting attention from individual AI systems to cross-institutional pathways \cite{SultanaAhmed2026CrossInstitutional}. We build on this perspective and extend it to datafied citizen services, where documents, records, and institutional decisions depend on connections between institutions even when AI is not involved. This motivates our third research question: How do individualistic institutional audits fail to ensure the quality of data valuation involving patching, and what adaptations do audit mechanisms need in theory and application to address data devaluation in citizen service ecosystems?

\section{Methods}

We conducted 41 semi-structured interviews between March and August 2026 with people who had experienced difficulties when moving documents, records, or other forms of information across institutions in Bangladesh. These experiences included transferring educational credentials, presenting medical records, proving professional skills, applying for financial services, and using records across government offices. Through these interviews, we examined how information that was valid in one institution could lose recognition or practical value in another institution. 

\begin{table}[t]
\centering
\renewcommand{\arraystretch}{0.85}
\setlength{\tabcolsep}{4pt}

\begin{tabular}{|c|}
\hline

\multicolumn{1}{|c|}{\textbf{Total Participants:} 41 (Female: 16, Male: 25)}\\
\hdashline

\multicolumn{1}{|c|}{\textbf{Age Range (in Years)}}\\
\hdashline
\multicolumn{1}{|c|}{
\begin{tabular}{r@{\hspace{8pt}}l}
18--25: & 7 (Female: 2, Male: 5)\\
26--35: & 14 (Female: 6, Male: 8)\\
36--45: & 10 (Female: 3, Male: 7)\\
46--55: & 7 (Female: 3, Male: 4)\\
56+: & 3 (Female: 2, Male: 1)
\end{tabular}
}\\
\hdashline

\multicolumn{1}{|c|}{\textbf{Education}}\\
\hdashline
\multicolumn{1}{|c|}{
\begin{tabular}{r@{\hspace{8pt}}l}
Primary Education or Below: & 3 (Female: 1, Male: 2)\\
Secondary School Certificate (SSC): & 8 (Female: 4, Male: 4)\\
Higher Secondary Certificate (HSC): & 10 (Female: 3, Male: 7)\\
Bachelor's Degree / Pursuing Bachelor's: & 16 (Female: 6, Male: 10)\\
Master's Degree or Above: & 4 (Female: 2, Male: 2)
\end{tabular}
}\\
\hdashline

\multicolumn{1}{|c|}{\textbf{Occupation}}\\
\hdashline
\multicolumn{1}{|c|}{
\begin{tabular}{r@{\hspace{8pt}}l}
Student: & 6 (Female: 2, Male: 4)\\
Salaried Employee: & 11 (Female: 3, Male: 8)\\
Self-employed: & 8 (Female: 2, Male: 6)\\
Skilled / Manual Worker: & 5 (Female: 1, Male: 4)\\
Homemaker: & 4 (Female: 4, Male: 0)\\
Not Currently Employed: & 4 (Female: 3, Male: 1)\\
Other / Retired: & 3 (Female: 1, Male: 2)
\end{tabular}
}\\
\hdashline

\multicolumn{1}{|c|}{\textbf{Domain(s) of Reported Experience}}\\
\multicolumn{1}{|c|}{\textbf{(multiple possible)}}\\
\hdashline
\multicolumn{1}{|c|}{
\begin{tabular}{r@{\hspace{8pt}}l}
Education: & 13 (Female: 3, Male: 10)\\
Employment \& Skills: & 14 (Female: 6, Male: 8)\\
Healthcare: & 10 (Female: 3, Male: 7)\\
Finance: & 11 (Female: 3, Male: 8)\\
Land \& Property: & 9 (Female: 4, Male: 5)\\
Public/Government Services: & 16 (Female: 7, Male: 9)
\end{tabular}
}\\

\hline
\end{tabular}

\caption{The demographic details of the participants}
\label{Tab:demo}
\end{table}

\subsection{Participant Recruitment}

We recruited participants through community networks, word of mouth, and snowball sampling. To participate in the study, individuals had to be at least 18 years old and have personal experience of moving a document or record between institutions. This included experiences in which a document was rejected, questioned, downgraded, or required additional verification before it could be used. After each interview, we asked participants to share information about the study with other people who might have similar experiences. Participants' experiences covered several institutional domains, including education, healthcare, professional and vocational skills, and financial services. Some participants also discussed experiences involving employment, land records, and public services. This range allowed us to examine whether similar problems appeared across different types of institutions. Participant details are summarized in Table~\ref{Tab:demo}.

\subsection{Interviews}

We conducted semi-structured interviews in Bengali, the native language of both the participants and the researchers. Interviews were conducted either in person or through Zoom, depending on participants' preferences and whether members of the research team were in Bangladesh at the time of the interview. Each interview lasted approximately 30 to 60 minutes and was scheduled according to the participant's availability. With participants' consent, 38 of the 41 interviews were audio-recorded. For the remaining three interviews, detailed interview notes were taken.

We first asked participants to describe the document or record, the institution that created it, and why they needed to use it elsewhere. We then asked what happened when they presented the record to the receiving institution. The interview guide covered five main areas: (1) how participants obtained the original document or record; (2) how they tried to use it in another institution; (3) why the receiving institution accepted, questioned, downgraded, or rejected it; (4) what additional work participants had to do to make the record usable; and (5) what personal, financial, or institutional consequences followed from the process. We also asked participants about both digital and non-digital interactions. These included online application systems, institutional databases, paper forms, official letters, in-person visits, phone calls, and communication between offices. Open-ended questions and flexible follow-up prompts encouraged participants to describe their experiences in detail. We asked them to explain the order of events, the institutions involved, the documents they were asked to provide, and the steps they took when their records were not accepted.

\subsection{Data Collection and Analysis}

All recorded interviews were transcribed in Bengali and then translated into English. Identifying information was removed before open coding and thematic analysis \cite{76boyatzis1998transforming, fereday2006demonstrating}. Two members of the research team repeatedly read the transcripts to become familiar with participants' experiences and to identify recurring patterns across the interviews.

We used an iterative process of open coding and thematic analysis. In the first round, we organized the data around the domains represented in the interviews, including education, employment and skills, healthcare, finance, land and property, and public/government services. Within these domains, we open coded participants' descriptions of what happened when information moved between institutions. The initial codes captured specific problems and actions described by participants. Examples included ``qualification not treated as equivalent,'' ``local medical report rejected,'' ``no way to verify the issuing institution,'' ``asked to collect the same document again,'' ``required to visit multiple offices,'' and ``payment needed to move a file.''

In the second round, we compared these codes across the different domains. Rather than treating each case as a separate problem in various fields, we examined the common processes that appeared across them. A pattern appeared across the different domains. Both institutions often functioned correctly according to their own rules, while the failure occurred in the relationship between them. We first grouped the codes based on the relationship between institutions. This identified three forms of data devaluation: missing connections, unequal recognition across existing connections, and connections that became obsolete over time. We then examined how participants responded to these problems, including the workarounds they used and the cases where those workarounds still failed.

\subsection{Positionality and Ethical Considerations}

This study was reviewed and approved by our institution's Institutional Review Board. Participation was voluntary, and informed consent was obtained before each interview. Participants were informed about the purpose of the research and their right to decline to answer any question or stop the interview. We removed participants’ names and other direct identifiers from the interview materials. To reduce re-identification risk, we omitted the names of organizations with which participants had employment, educational, or training affiliations. 

All authors are Bangladeshi and are native Bengali speakers. We were born and raised in Bangladesh and are familiar with the country's cultural, linguistic, and institutional contexts. This shared background helped us communicate with participants in their preferred language and understand references to documents, offices, social relationships, and bureaucratic practices that might be difficult to explain to an outsider. It also helped participants speak openly about frustrating or sensitive institutional experiences. At the same time, we recognize that our positions as university researchers differ from the positions of many participants. Our familiarity with the context could also lead us to assume that we already understood an experience. To reduce this risk, we used follow-up questions and asked participants to explain events in their own words. During analysis, we returned to the interview transcripts and notes and discussed alternative interpretations within the research team rather than relying only on our prior knowledge.
\section{Findings}

\begin{wrapfigure}{r}{0.55\textwidth}
\vspace{-15pt}
\centering
\includegraphics[width=0.98\linewidth]{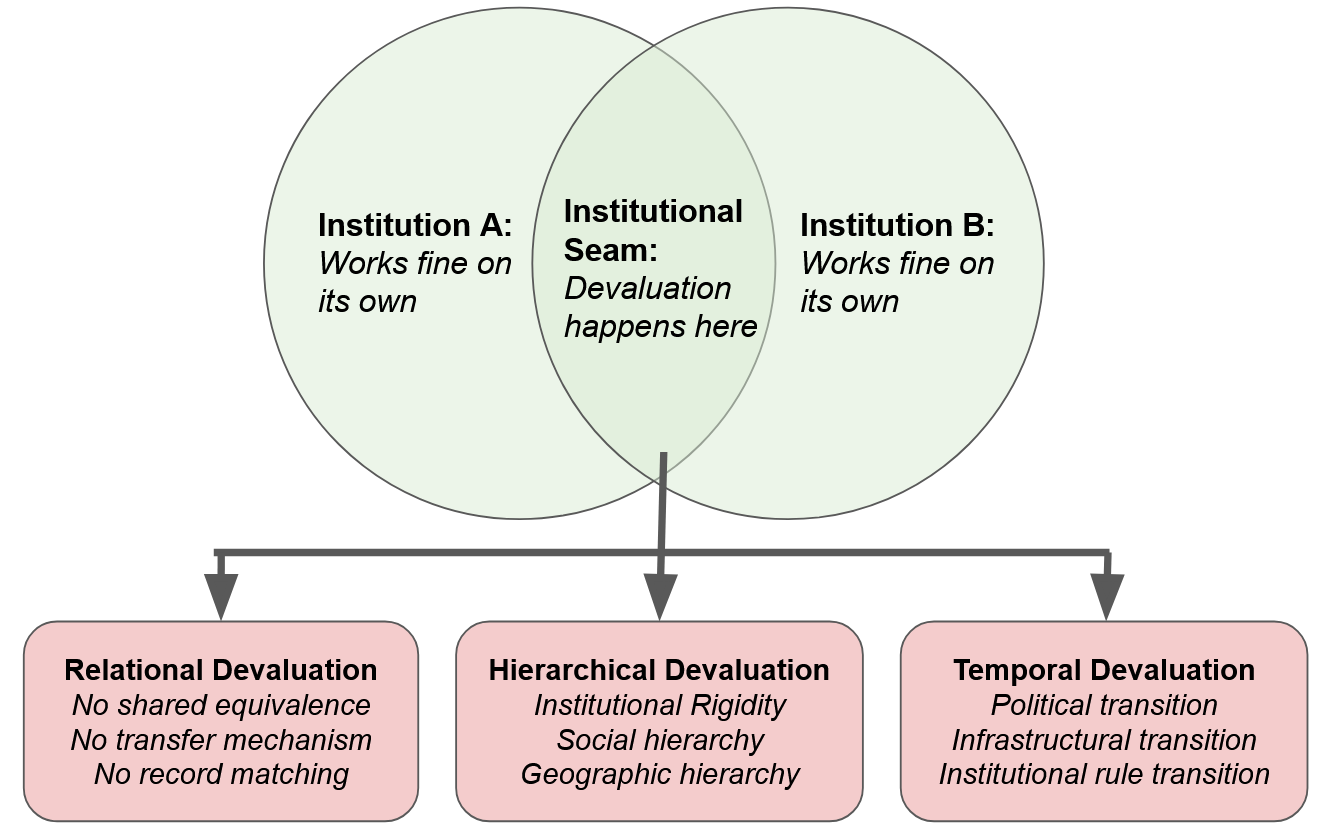}
\vspace{-5pt}
\caption{Where devaluation emerges across institutional seams. Our findings show that two institutions can function properly on their own while problems emerge when their records, procedures, or decisions need to connect. We identify three forms of cross-institutional data devaluation: relational devaluation, where the necessary connection is missing; hierarchical devaluation, where the connection exists but recognition is unequal; and temporal devaluation, where a previously working connection becomes obsolete after institutional change.}
\label{fig:institutional-seam-breakdowns}
\vspace{-15pt}
\end{wrapfigure}

In this section, we present our findings. The first three themes identify three ways cross-institutional data devaluation emerges. We refer to these three forms as relational devaluation, hierarchical devaluation, and temporal devaluation, as shown in Figure~\ref{fig:institutional-seam-breakdowns}. The fourth theme shows the citizen-borne workarounds people use to make these failed connections work. The final theme identifies the limits of these workarounds and shows why, in some cases, people still cannot make the institutional connection work.

\subsection{Missing Links Between Standalone Systems}

In this theme, we show cases where the necessary connection between two institutions had not been established. Because this link was missing, records that worked within one institution could not move, match, or be treated as equivalent in another. We found three forms of this devaluation. First, the institutions had no mechanism for transferring a recognized record from one institution to another. Second, they had no common basis for treating different forms of evidence as equivalent. Finally, they had no mechanism for matching two records to the same person or case.

\subsubsection{No Transfer Channel Between Systems}

Five participants described cases where information, records, or institutional value could not move directly from one institution to another because there was no accepted procedure for transferring it across the seam. P11 experienced this when he tried to move from an Alia madrasa to a general school after completing Class 8.

\begin{quote}

\textit{``The school told us that my madrasa education was officially recognized as equivalent to Class 8. However, their online admission system had no option for entering a madrasa student as a transfer applicant, and there was no transfer form or office that could process my case. They told me that instead I would have to take an additional admission test before they could consider me for Class 9." \textbf{(P11)}}

\end{quote}

The state had already recognized the two grade levels as equivalent, and the school also accepted this. The problem appeared when P11's recognized madrasa record had to enter the general school's admission system. P35 experienced a similar problem when trying to move money between two banks.

\begin{quote}
\textit{``I needed to transfer funds from my account at Citizens Bank to a new account at Janata Bank. When I presented my bank cheque and transfer documents at the Janata Bank branch, they refused to process it. They told me they did not have any mechanism for transferring money like this because Citizens Bank was newly established. To move my money, I had to withdraw the cash from Citizens Bank, take it to Janata Bank, and deposit it manually" \textbf{(P35)}}
\end{quote}

P35’s account and transaction information was valid in Citizens Bank, but Janata Bank could not use it because there was no transfer pathway between the two banks. His information therefore remained valid in Citizens Bank but lost its practical value when he tried to use it at Janata Bank. 

Together, these cases show that having valid information or value within one institutional system does not guarantee that it can move into another. In both cases, the devaluation occurred at the transfer between institutions.

\subsubsection{No Basis for Cross-Institutional Equivalence}

Seven participants described cases where two institutions accepted different forms of evidence and had no way to treat them as equivalent. A person's experience, history, or qualification could be valid in one institution, but the other institution required a different form of evidence. As a result, participants could not use one form of evidence in place of the other.

P13 had worked as a plumber for almost 15 years and had an experience certificate from a private company. However, a government plumbing position required a Bangladesh Technical Education Board (BTEB) trade certificate.

\begin{quote}
\textit{``When I applied for a government plumbing position, the office told me that work experience and a company certificate were not equivalent to a BTEB trade certificate. They said that my form of experience could not be treated as the qualification required for the position." \textbf{(P13)}}
\end{quote}

P13’s work experience was still valid and recognized by his employer. However, it lost practical value in the government recruitment process because there was no institutional mechanism for translating that experience into the type of qualification the government accepted. A similar lack of equivalence appeared between microfinance and formal banking. P24 had borrowed from BRAC NGO for several years and had always repaid his loans on time. He later approached BRAC Bank for an education loan.

\begin{quote}

\textit{``The loan officer asked about my income, assets, and borrowing history. I explained that I had taken loans from BRAC NGO for years and always repaid on schedule. However, the bank said they would not consider my NGO loan history." \textbf{(P24)}}

\end{quote}

BRAC NGO had several years of records showing that P24 repaid his loans on time. However, BRAC Bank did not treat this history as relevant evidence for lending process. Together, these cases show how institutional connections can fail when institutions do not treat different forms of recognized evidence as equivalent. In both cases, the existing evidence was valid, but the other institution required a different form of proof.

\subsubsection{No Cross-System Record Linkage}

Four participants described cases where two institutions held valid records about the same person, but there was no established mechanism for matching those records across their systems. Small differences in names, IDs, or other details could prevent the systems from recognizing that the records referred to the same person. P17 faced this when a government recruitment board compared his National ID card with his university certificate. In Bangladesh, ``Md." and ``Mohammad" are two written forms of the same name, and different offices use different forms.

\begin{quote}
\textit{``My NID has my name as Mohammad Abul Kalam [pseudonym] and my certificate has it as Md. Abul Kalam. The board flagged my application because the two names did not match exactly. My university then gave me a letter confirming that both names belong to the same student with the same roll number." \textbf{(P17)}}
\end{quote}

Neither the university nor the NID system had produced an incorrect record. However, the difference in how P17's name was recorded created a problem when the recruitment board tried to connect the two records. P17 therefore had to obtain additional confirmation from the university to show that both records referred to the same person.

\subsection{Unequal Recognition Across Existing Couplings}

In this theme, we discuss cases where a connection between two institutions already existed. However, due to hierarchical differences, the institutions did not connect equally in practice. These hierarchies could come from how rigid the institution's rules were, the social position of the person carrying the record, or the geographic location from which the record came.

\subsubsection{Rigid Recognition Based on Institutional Status}

Five participants described cases where one institution followed more rigid rules and treated the institution that produced the record as less trustworthy. In these cases, the coupling between the institutions already existed. However, one institution’s unbending rules became a proxy for judging the documents. P31 experienced this while applying for a government position at the Dhaka Power Distribution Company (DPDC). He was a job applicant and had never worked at DPDC. 

\begin{quote}
\textit{``During the interview, the panel looked at my certificate and asked me, `Which university did you graduate from?' They then said, `We have never even heard the name of this university before and we may not consider you for the position.' Although my university was UGC-approved, they looked down on me because of where I studied." \textbf{(P31)}}
\end{quote}

P31’s university had issued an officially recognized degree, and DPDC had accepted his application and invited him for an interview. However, the panel applied a more rigid judgment to his degree because it came from a university they treated as lower status. A similar problem appeared when participants moved between schools with different levels of institutional rigidity. P34 shared his experience after completing Class 7 and applying to several schools with more rigid admission rules.

\begin{quote}
\textit{``After looking at my transfer certificate, they told me that they would not accept students from schools like mine. They told me to apply for Class 6 or 7 if I wanted a chance to enroll. They did not want someone from a school like mine to be admitted and later hurt the school's reputation by doing bad in the exam." \textbf{(P34)}}
\end{quote}

P34's transfer certificate showed that he had completed Class 7, but the schools did not give it the same value they might have given to a certificate from a school they treated as higher status. Their judgment was based on the status of his previous school rather than whether the certificate itself was valid. In both cases, the records remained formally valid, but their value was reduced because of rigid recognition based on the status of the institution that produced them.

\subsubsection{Social Hierarchies of Recognition}

Four participants described cases where a connection between institutions existed, but the social position of the person carrying the records affected whether that connection worked. When the records were presented by people with less social influence or authority, the institution was less willing to act on them. As a result, the same institutional connection could work for some people while records carried by others lose recognition. P26 experienced this after her father died and she went to the land office to divide his property among his three daughters.

\begin{quote}
\textit{``I took all the necessary documents, like proof of inheritance, ownership documents, and my National ID card (NID), and went to the office myself. But they told me that I needed to bring a brother, an uncle, or some other male family member. I told them that I do not have a brother and do not want to involve other male relatives. But they said they needed a man for this kind of land document work and would not process it without one." \textbf{(P26)}}
\end{quote}

The documents, such as proof of inheritance, ownership documents, and the National ID card, were all valid. Different offices had issued these documents, and the land office could read and understand them. However, the connection between these records and the land office did not work when P26 presented them herself. This case shows how social hierarchy can determine whether an existing connection between institutions actually works.

\subsubsection{Geographic Hierarchies of Trust}

Five participants described cases where official records and applications were treated differently depending on where the record was produced or where the person came from. Institutions sometimes applied more rigid requirements to records from district-level or local institutions, or to people living in particular geographic areas. P9 experienced this after her husband died in Moulvibazar, a district outside the capital.

\begin{quote}
\textit{``I submitted all the required documents, including the death certificate, but the insurance company did not trust it. They told me that because the death occurred outside Dhaka, they needed additional verification and supporting records. I then spent more than two years collecting documents, visiting offices, and repeatedly proving that my husband had actually died." \textbf{(P9)}}
\end{quote}

The hospital had officially recorded the death and issued the certificate. The insurance company also had a process for receiving the certificate and assessing the claim. However, the certificate was subjected to more rigid requirements because it came from outside Dhaka. A similar hierarchy appeared in passport applications. P36 described this problem for people living in Cox's Bazar and Teknaf.

\begin{quote}

\textit{``Because people sometimes cross the international border through these areas, our applications go through much more verification. We have to provide bank documents and many additional records, including information about our family properties and the NIDs and property records of our parents and grandparents. Someone in Dhaka may get a passport within a month, while for us it can take six months to a year." \textbf{(P36)}}

\end{quote}

P36’s identity and supporting records were valid, but they were not treated as enough proof in the passport process because of where he lived. In both cases, geography led institutions to apply more rigid requirements before they were willing to act.



\subsection{Institutional Coupling Becomes Obsolete}

In this theme, we discuss cases where a connection between the institutions already existed and had worked before. Over time, that coupling became obsolete because of political change, digitization, or changes in the rules about which records an institution would accept. 

\subsubsection{Recognition Changed After Political Transition}

Six participants described cases where records lost recognition after a political transition. In these cases, the institution itself remained the same, but the new administration changed how it treated records issued under the previous administration. P41 described one such case.

\begin{quote}
\textit{``I completed an electrical training program organized by the Union Parishad and received a certificate from the same office. But after the administration changed in 2025, the new officials said they would not recognize certificates issued under the previous administration. They told me that I would have to take the matter to a higher authority to get it resolved." \textbf{(P41)}}
\end{quote}

The certificate itself did not change. What changed was the willingness of the issuing office to stand behind it. This shows how a political transition can break a connection that had previously worked.


\begin{wrapfigure}{r}{0.58\textwidth}
\vspace{-15pt}
\centering
\includegraphics[width=0.98\linewidth]{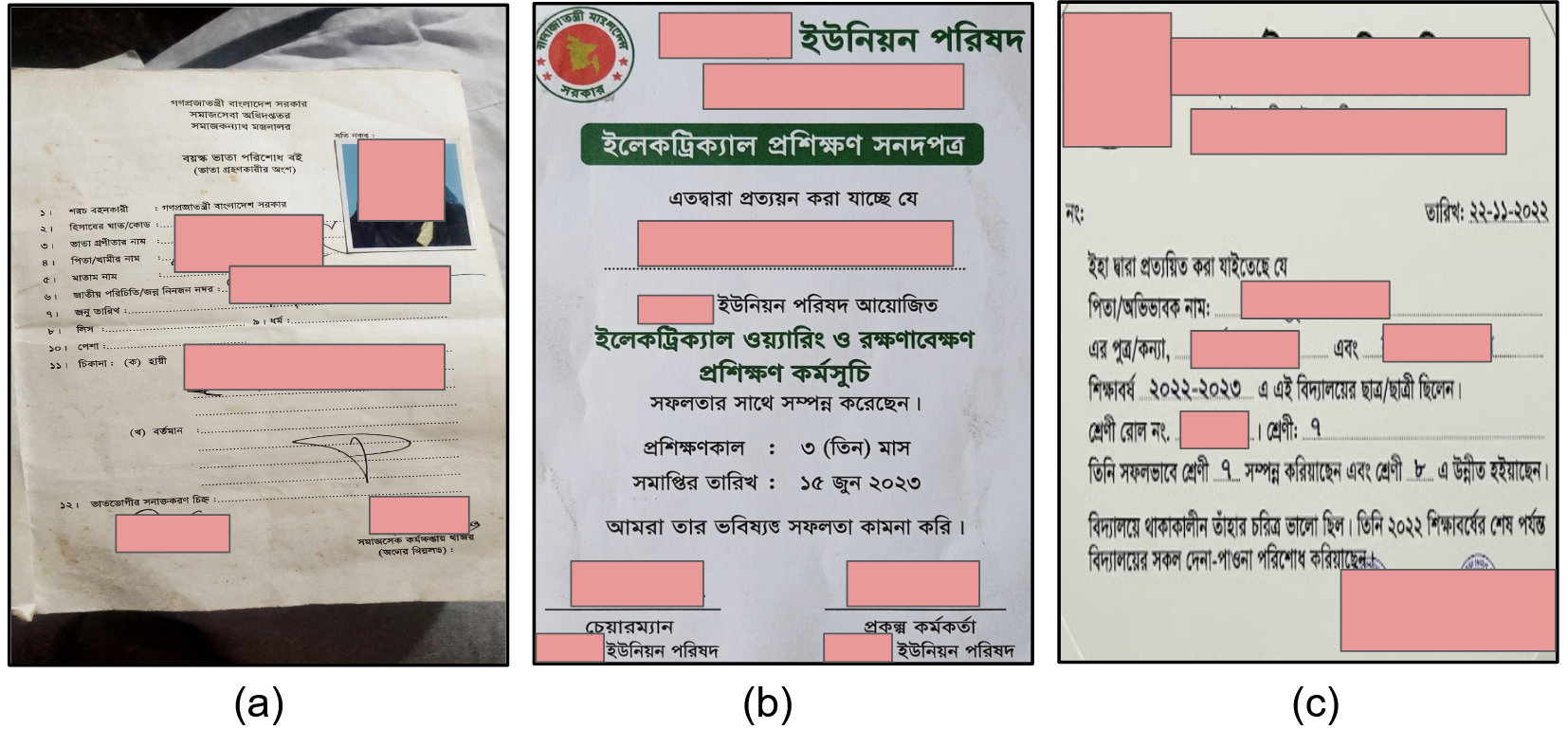}
\vspace{-15pt}
\caption{Examples of records that remained valid but lost practical value across institutional seams. (a) An old-age allowance record became unusable after payments moved to mobile banking because the holder's record was missing from the online system and her NID and SIM information did not match. (b) An electrical training certificate lost verifiability after a government transition, when new officials refused to verify certificates signed by the previous administration. (c) A school transfer certificate was recognized but given less value because it came from a lower-status school and was subjected to more rigid admission rules, preventing admission to Class 8.}
\label{fig:record-examples}
\vspace{-15pt}
\end{wrapfigure}

\subsubsection{Compatibility Broken by Infrastructural Transition}

Seven participants described cases where an existing institutional coupling was disrupted after a change in infrastructure. A process that had worked before stopped working when institutions introduced new technologies, systems, procedures, or data standards. P21 described how this affected his grandmother's old-age allowance. 

\begin{quote}
\textit{``My grandmother received her old age allowance without problems for years. But when payments were digitized and shifted to mobile banking, it stopped. At the Upazila Social Services Office, we were told her record was missing from the online system and her National ID and SIM information did not match. Her allowance stayed suspended for months while officials repeatedly said the data was being corrected, a request had been sent to Dhaka, or a system update would fix it." \textbf{(P21)}}
\end{quote}

The allowance system had worked for years, but the connection broke when her older records had to move into the new system. This case shows how an infrastructural transition can break a connection that had previously worked.

\subsubsection{Compatibility Broken by Institutional Rule Changes}
Three participants described cases where an institutional coupling had worked before but later stopped working after the institution changed its rules. Before the change, a record from a particular institution was accepted without difficulty. After the change, the same kind of record from the same institution was no longer accepted. P15 experienced this when renewing his father's driving licence.

\begin{quote}
\textit{``For years my father renewed his driving licence using a medical fitness certificate from our District Civil Surgeon's office. Later, BRTA changed its rules and said it would only take medical reports from its own designated centers or from certain government hospitals in Dhaka. When I went to renew his licence, the officer refused our Civil Surgeon's report. The report was genuine, and the Civil Surgeon is the highest government medical authority in our district, but they said it was no longer on their accepted list." \textbf{(P15)}}
\end{quote}

The Civil Surgeon's report had been accepted for the same process for years, and the office that issued it remained a government medical authority. The problem appeared after BRTA changed its rule about which medical institutions it would recognize. Recognition that had continued for years was lost after the receiving institution changed what it was willing to accept. Figure~\ref{fig:record-examples} shows examples of records from our findings (4.1-4.3) that remained valid but lost practical value across institutional seams.

\subsection{Citizen-Borne Workarounds for Interoperability Gaps}

The first three themes showed that two institutions may work properly on their own but still face problems when they need to connect. In this theme, we focus on the workarounds participants used to make these connections work. They visited offices, collected new evidence, contacted intermediaries, repeated procedures, and sometimes paid gatekeepers. These citizen-borne workarounds can move an individual case forward, but they do not create a lasting connection between the institutions. The same problem may happen again for the next person.

\subsubsection{Paying Gatekeepers to Activate Formal Pathways}

Fourteen participants described cases where paying gatekeepers or intermediaries helped make the connection between institutions. In these cases, a formal connection already existed, but it did not work in practice until participants paid gatekeepers or intermediaries. P2 faced this problem while applying for a passport. 

\begin{quote}
\textit{``I submitted all the required documents, including one issued by another government office but the officer repeatedly delayed the verification and asked me to return several times. Eventually, an intermediary told me it would not move forward unless I paid extra. I paid around BDT. 2500 (USD. 20), and soon afterward the passport office confirmed the same document and processed my application." \textbf{(P2)}}
\end{quote}

This example shows how paying gatekeepers worked as a workaround to make the connection between two institutions work.  A similar form of problem appeared when P10 applied for an electricity connection for his electrical shop in Chittagong.

\begin{quote}
\textit{``The electricity office was waiting for confirmation from the office that issued my trade license, but the file was not being released. The process stayed delayed for almost a month. Then someone told me that if I paid some money, the file problem would be solved. I paid around BDT. 4000 (USD. 35), and the process moved forward." \textbf{(P10)}}
\end{quote}


Together, these cases show that a formal connection between institutions could exist but still not work. In both cases, participants had already provided the required documents, but they had to pay extra money to make the connection work.

\subsubsection{Reproducing Evidence Inside the Receiving Institution}

Five participants described cases where they reproduced existing evidence as a workaround to make the connection between institutions work. Participants chose to repeat tests or procedures and create the same evidence again within the second institution so that their case could move forward. P25 experienced this after visiting his local Hospital with severe fatigue, leg swelling, and loss of appetite. A doctor there ordered several tests, including Serum Creatinine, Urea, Urine R/E, and CBC, and then referred him to a nephrologist at Apollo Hospital.

\begin{quote}
\textit{``When I showed the nephrologist my reports, he told me that they would not accept medical reports from another hospital. I needed to continue my treatment, so I decided to do the same tests again at Apollo. I spent nearly BDT. 30,000 (USD. 250), and the results were almost the same as the reports I already had." \textbf{(P25)}}
\end{quote}

P25 already had medical evidence from the first hospital, but Apollo would not use those reports for his treatment. So, he chose to reproduce the same evidence inside Apollo so that his treatment could move forward. A similar workaround appeared when P7 applied for a labor visa to Saudi Arabia. He submitted the required documents through a recruitment agency, including proof of Hepatitis B vaccination. He had already received the vaccine at his local Upazila Health Complex and had a vaccination card.

\begin{quote}
\textit{``But the agency said they would only accept vaccination records from a particular list of hospitals in Dhaka. I thought that if I took the vaccine again at one of those hospitals they would accept the report and move my visa process forward. So I decided to take the vaccine again and submit the new card." \textbf{(P7)}}
\end{quote}

P7's original vaccination record was valid, but the visa agency only accepted records from particular hospitals. He therefore chose to get the same evidence from an accepted hospital. Together, these cases show how participants reproduced evidence as a workaround when existing evidence could not move across institutional boundaries.

\subsubsection{Manual Integration of Fragmented Records}

Six participants described cases where they manually combined records from different institutions as a workaround to make separate institutions connect. In these cases, information about the same person, property, or event was spread across different institutions, but those records were not connected to one another. Participants therefore collected the separate records themselves and combined them into one form of proof so that their case could move forward.

P39 experienced this when his family divided their ancestral land. The division was documented through legally registered deeds. However, registering the deeds did not automatically update the mutation records, khatian, land-tax records, and other records held by different government offices.

\begin{quote}
\textit{``Years later, when part of the land was acquired for a government project, we submitted the deeds to claim compensation. The land office said the deeds were not enough because our names had not been updated in the other records. We then collected mutation documents, khatian records, tax receipts, and other evidence from several offices and combined them as proof of ownership." \textbf{(P39)}}
\end{quote}

In P39's case, the ownership information existed across several government record systems, but those systems did not connect the records themselves. P39's family therefore collected the separate records and combined them into one form of proof as a workaround. By doing this, they manually created the connection between record systems that remained institutionally separate.

\subsection{Limits of Citizen-Borne Workarounds}

The previous theme showed how participants tried to make failed institutional connections work through their own labor. However, in some cases participants tried similar workarounds but still could not make the connection work. In these cases, their own workarounds were not enough to fix the connection.

\subsubsection{When Institutions Could Not Agree on a Shared Verification Standard}

Seven participants described cases where each institution followed its own verification standard but no shared standard existed between them. Participants tried to solve this problem by collecting additional confirmation, carrying records between institutions, changing the form of the record, or offering informal payments. However, these workarounds failed because they could not create a standard of verification that both institutions would accept. P6 experienced this when he tried to use a computer training certificate for a government job. The training center's standard verification process was to provide an official verification letter, while the government office required verification through a government portal.

\begin{quote}
\textit{``I asked the training center to verify my certificate. They told me their standard procedure was to provide an official letter confirming that the certificate was theirs and that I had completed the course. I took that letter to the government office, but they said their standard required verification through a government portal. The training center did not use that portal, so I could not use the certificate for my application." \textbf{(P6)}}
\end{quote}

P6 tried to solve this himself by returning to the training center, collecting an additional verification letter, and carrying it back to the government office. However, the government office required verification through a portal that the training center could not provide. Each institution followed its own verification standard, and P6 could not make the two meet. A similar problem appeared when P27 tried to use an employment record to prove his employment history for a bank loan.

\begin{quote}
\textit{``My former company only issues digital, QR-code-stamped PDFs, but the bank required a paper document with a physical stamp and signature. I went back to HR, and they printed and stamped the document, but the bank rejected it again because it did not have an original physical signature. I was going back and forth with different versions of the same document, but neither side would accept the other's format." \textbf{(P27)}}
\end{quote}

P27 tried to make the two standards work together by asking his former company to change the form of the document. Even after the company printed and stamped it, the bank still rejected it because its standard required an original physical signature. Together, these cases show a limit of citizen-borne workarounds. Participants could collect more evidence, obtain additional confirmation, and carry different versions of the same record between institutions. However, these efforts could not create the shared verification standard that was missing between the institutions.

\subsubsection{When No Actor Was Responsible for the Crossing}

Four participants described cases where both institutions had officials responsible for matters within their own systems, but no one was responsible for the connection between them. Participants tried to make the connection work by carrying records between institutions and asking each office to act. However, these workarounds failed because neither office had the authority to complete the handover. P28 experienced this after his family's area was transferred from a Union Parishad to a municipality.

\begin{quote}
\textit{``Our area was previously under the Union Parishad, where my father paid holding tax for many years. The municipality accepted that the old Union Parishad receipts were genuine, but said the corresponding records had never been transferred into its files. The Union Parishad also said it no longer had authority to update them because our area was now under the municipality. So we kind of moved from one office to the other, but neither office actually helped us solve the problem." \textbf{(P28)}}
\end{quote}

P28's family carried the original receipts, visited both offices repeatedly, and asked each one to act. The municipality could not act because it did not hold the records, while the Union Parishad could not act because it no longer had authority. What was missing was an actor with authority over the handover itself.

\subsubsection{When the Update Path Could Not Be Restored}

Five participants described cases where a record created under an older system could no longer be moved into the system that replaced it. Participants went to the responsible offices, brought supporting records, and asked officials to update or restore the older record. However, these workarounds failed because the route between the old and the new system had already been closed. P40 experienced this when he tried to update his 13-digit birth certificate to the newer format required for his passport application.

\begin{quote}
\textit{``I had never needed to change it before because the 13-digit certificate had worked everywhere. Now I have been trying for a long time to update it, but they have not been able to do it. Even at the birth registration office, they cannot properly find my old record because my birth certificate has only 13 digits." \textbf{(P40)}}
\end{quote}

P40 returned to the birth registration office to convert his older certificate into the newer format. However, there was no longer a working route for connecting the older record to the current system. A similar problem appeared when P32 tried to restore his father's Freedom Fighter record. His father had used a handwritten certificate to receive his allowance, but after the system moved online, his name was missed because the spelling on the certificate did not match his NID.

\begin{quote}
\textit{``I went to the Muktijoddha Sansad office and the DC office many times with his original documents, old press clippings, and past allowance receipts. The officials checked everything and said the documents were real, but the online entry portal for old manual records was permanently closed." \textbf{(P32)}}
\end{quote}

Together, these cases show a limit of citizen-borne workarounds. Participants could gather evidence that officials accepted as genuine, but genuineness was not the problem. The route through which an older record could enter the current system had been closed, and no amount of evidence could reopen it.

\section{Discussion}
Our interview study investigated cross-institutional data devaluation and found how valid records can lose recognition, credibility, or practical value when they move across institutional seams in datafied citizen services. Our findings on the forms of data devaluation, citizens' workarounds of patching mechanisms, and the failure of patching open discussion on both theory and design fronts.

\subsection{How Valid Records Lose Value at Institutional Seams}

Our findings show that in citizen services that involve multiple institutions, a record can remain accurate, complete, and officially valid while losing its ability to produce action. Across the forms of devaluation identified in Sections 4.1--4.3, we found three conditions of the institutional relationship through which this devaluation occurs: the connection may be missing, unequal, or become obsolete over time. Across these conditions, institutions could follow their own protocols while the record still failed at the seam between them.

\subsubsection{Recognition Does Not Always Carry Across Institutions} Bowker and Star show that institutions use classifications and categories to organize people and experiences, and when something does not fit those categories, the institution may fail to recognize it \cite{BowkerStar1999,StarBowker2007}. In our cases, however, participants were often clearly recognized within the systems they came from. P13 was recognized as an experienced plumber by his employer, but the government office did not treat that experience as equivalent to the required BTEB qualification. In P11's case his madrasa education was already recognized as equivalent to Class 8, but there was no transfer pathway for using that record in the school's admission process. These cases show that recognition within one institution does not ensure that a record can be used in another institution.

\subsubsection{Valid Records Can Still Be Valued Differently} Documents also gain authority through institutional forms that connect them to their issuing authority \cite{Hull2012,SharmaGupta2006,Porter1995}. Similarly, valuation research shows how institutions decide which forms of evidence are more credible than others \cite{EspelandStevens1998,Lamont2012}. Our findings show that even when a record was formally valid and recognized, its value could still depend on the institution that produced it. DPDC recognized P31's UGC-approved university degree, and the school recognized P34's transfer certificate. Yet both records were given less value because of the status of the institutions that produced them. Across this theme, institutional status, geography, and social position could change how much value a formally valid record received.

\subsubsection{Institutional Change Can Make Valid Records Unusable} Singh and Jackson use the concept of resolution to show that a person may be clearly represented in one system but become low-resolution when their record enters another institutional process \cite{SinghJackson2021}. Our findings show that even when the record holder does not move, the institutional environment around the record can change instead. P15's father's medical fitness certificate had been accepted for years but became unusable after BRTA changed which medical institutions it would recognize. Similarly, P41's certificate remained the same after a political transition, but the issuing office would no longer verify it. A record could therefore lose value even while remaining in the hands of the same person and within the same national setting.

\subsubsection{Proximity Does Not Guarantee Equal Recognition} This also extends recent HCI work on data migration. Rohanifar et al. show how immigrants' records can lose value when they move across national and institutional regimes \cite{RohanifarEtAl2026}. In our cases, cross-institutional data devaluation appeared between institutions operating within a single country and even between institutions under shared organizational ownership. P24's BRAC NGO repayment history was not recognized by BRAC Bank, even though both belonged to the same parent organization. P35 similarly could not transfer money directly from Citizens Bank to Janata Bank because no transfer mechanism had yet been established between the two banks. These cases show that neither national proximity nor shared organizational ownership alone guarantees that records or institutional value can move across institutions.

In many cases, people or institutions with less power depended on institutions with greater authority to recognize and act on their records. Prior work has shown that standards and classifications carry political choices about what becomes comparable, credible, and valuable \cite{BowkerStar1999,EspelandStevens1998,Lamont2012}. Our findings point to the other side of this argument, where missing pathways, unequal recognition, and connections that become obsolete over time can also have political consequences. 


\subsection{When the Citizen Becomes the Patch}

Our findings showed how participants used workarounds to make failed institutional connections work so that their intended citizen services could continue. They paid gatekeepers to move existing pathways forward, reproduced evidence inside another institution, and manually combined records from different institutions. In each case, participants provided labor or coordination that the institutions did not provide. These workarounds could make a particular case move forward, but they did not create a lasting connection between the institutions.

\subsubsection{Citizen-Borne Coordination Work} Prior work on \textit{articulation work} helps explain part of this labor. Articulation work is the coordination needed to bring separate activities together and keep a larger process working \cite{SchmidtSimone1996}. This work often remains invisible in formal accounts of organizations \cite{StarStrauss1999}. Prior work often describes such coordination as being done by workers or intermediaries \cite{SchmidtSimone1996,StarStrauss1999}. In our cases, however, the person seeking the service often had to do this work. For example, P25 carried his medical reports between hospitals and repeated tests so that his treatment could continue. 

Intermediaries also appeared in some of these workarounds. Prior work shows how brokers, operators, family members, and local helpers can help people connect with institutional systems \cite{Chaudhuri2019,CarswellDeNeve2021,RohanifarEtAl2026, coles2019accessing}. P2's case was different. A formal passport-verification pathway already existed, and he had already submitted the required documents. After he paid an intermediary, no new document or evidence was added. The same verification simply moved forward. In this case, the person had to bear an additional cost to make an institutional connection that was already supposed to work.

\subsubsection{From Governance by Patching to Data Patching} These workarounds both cohere with and differ from Veeraraghavan's concept of \textit{patching} \cite{veeraraghavan2022patching}. They cohere with patching in two ways. First, both respond to a specific breakdown rather than trying to redesign the whole system. Second, both use local information about where a process is failing to find a focused way to make it work. However, they differ in who performs the patch and what that actor has the authority to change. In Veeraraghavan's account, patching depends on an actor with jurisdiction who can change rules, procedures, or institutional arrangements. In our cases, the people performing the workaround had no such jurisdiction over either institution or the connection between them. They responded to the breakdown themselves by paying, reproducing evidence, or manually connecting records.

We call this citizen-borne form of patching \textit{data patching}. A data patch makes an institutional connection work for a particular case without repairing the underlying seam. The citizen carrying the record therefore becomes the patch. P2's payment moved his passport application, but it did not change the verification process for the next applicant. Similarly, P25's repeated tests allowed his treatment to continue, but they did not make reports from the first hospital acceptable for later patients. Data patching therefore shifts the work of making institutions connect onto the person while leaving the institutional relationship unchanged.

\subsubsection{When Successful Patching Hides Institutional Failure} Successful data patching can also hide the underlying institutional failure. Once P2's passport was processed or P25's treatment continued, the final outcome could make the process appear to have worked. What the outcome does not show is the informal payment, repeated procedures, travel, delay, or personal labor required to make the connection work. People therefore bear the cost of these seams even when they eventually receive the intended service. A successful individual outcome does not necessarily mean that the institutional seam has been fixed.

\subsection{Why Individualistic Institutional Audits Can Miss the Seam}

Our findings also showed the limits of citizen-borne data patching. Participants could not create a shared verification standard between institutions, create an actor with authority over the crossing, or restore an update path that no longer existed. We argue that these same limits also show why individualistic institutional audits can miss failures at institutional seams in citizen services that involve multiple institutions. If Institution A and Institution B are audited separately, both may appear to follow their own rules and procedures while the connection between them still fails. We show this through the same three conditions identified in Section 4.5.

\subsubsection{No Shared Standard to Measure Against}

Individualistic institutional audits usually need a standard against which a system or institution can be evaluated \cite{GrovesEtAl2024,WrightEtAl2024}. Section 4.5.1 showed that this becomes difficult when each institution has its own standard but no shared standard exists between them. P6's training center verified certificates through an official letter, while the government office required verification through a government portal. P27 faced a similar problem when his former company treated a QR-coded digital record as official, while the bank required an original physical signature. If these institutions are audited separately, each audit can examine whether the institution follows its own standard. The training center may correctly follow its verification procedure, and the government office may also correctly follow its own requirement. The same can be true for P27's former company and the bank. However, the record can still fail when it moves between them. An audit bounded within one institution can therefore find that its standard is being followed while leaving the missing standard between institutions unexamined. In these cases, the seam itself needs to be examined as part of the audit.

\subsubsection{No Accountable Actor at the Seam}

Audits also depend on identifying an actor who can explain or respond to a problem \cite{WilsonEtAl2021,YoungEtAl2022,RajiBuolamwini2019}. Section 4.5.2 showed that participants sometimes reached a point where both institutions had responsible actors within their own boundaries, but no actor had authority over the crossing between them. P28's case shows this clearly. The municipality accepted that the old Union Parishad records were genuine, but those records had never been transferred into its files. The Union Parishad still had the older records but no longer had authority to update them. The problem was therefore not that either institution had no responsible officials. What was missing was an actor with authority over the handover itself. An audit of the municipality could identify who was responsible for records within the municipality. An audit of the Union Parishad could do the same within that institution. However, auditing them separately would still leave the transfer between them without a responsible actor. This creates an accountability gap at the seam. Identifying responsible actors inside institutions is therefore not enough when no one has authority over the institutional crossing itself.

\subsubsection{No Continuity Over Time}

Auditing also becomes more difficult when systems and institutional arrangements change over time \cite{AngwinTobinVarner2017,AliEtAl2019}. Section 4.5.3 showed that participants sometimes held records that had been valid and usable under an earlier system, but the route for moving those records into the current system no longer existed. P40 faced this with his older birth certificate, while P32 faced a similar problem after the Freedom Fighter allowance system became digital. These cases create a different problem for individualistic institutional audits. An audit of the earlier system could show that the record was valid and usable under the rules that existed at that time. An audit of the current system could examine whether the present system follows its current rules. Yet the record can still become unusable because the connection between the earlier and current arrangements has disappeared. The failure therefore lies in the loss of continuity across institutional change. Looking only at the earlier system or only at the current system does not show whether records created under one arrangement can still remain actionable under the next. 

Taken together, these cases show a common problem. Individualistic institutional audits can examine Institution A and Institution B and still miss a failure that exists in the relationship between them. One institution may have its own standard, responsible actors, and functioning procedures, and the other may have the same, while the record still fails at the crossing. The same problem can occur over time when an earlier and a current system each work according to their own rules but the route between them disappears. The institutional seam therefore needs to become an object of audit and accountability. This motivates the Citizen-centered Cross-institutional Data Audit (CCDA) that we develop in the next section.


\subsection{Toward Citizen-centered Cross-institutional Data Audit (CCDA)}

Building on the limitations identified in Section 5.3 and the broader patterns across our findings, we propose the \textit{Citizen-centered Cross-institutional Data Audit (CCDA)}. Drawing on Cross-Institutional AI Auditing (CIAA) \cite{SultanaAhmed2026CrossInstitutional} sensitivities, CCDA is developed as an approach for datafied citizen service ecosystems where records, decisions, and services depend on connections between multiple institutions. The CIAA framework shifts attention from individual AI systems to the institutional pathways through which data moves \cite{SultanaAhmed2026CrossInstitutional}. Therefore, CCDA makes the institutional seam itself an object of audit rather than auditing each institution separately. Hence, CCDA is an extension of CIAA to datafied citizen services and will be useful for the government, policy makers, and the public-facing service design industry to audit across institutions and ensure quality citizen services. CCDA framework comprises five interrelated principles (see Figure ~\ref{fig:five-principles}), described below:



\begin{wrapfigure}{r}{0.48\textwidth}
    \vspace{-15pt}
    \centering
    \includegraphics[width=0.99\linewidth]{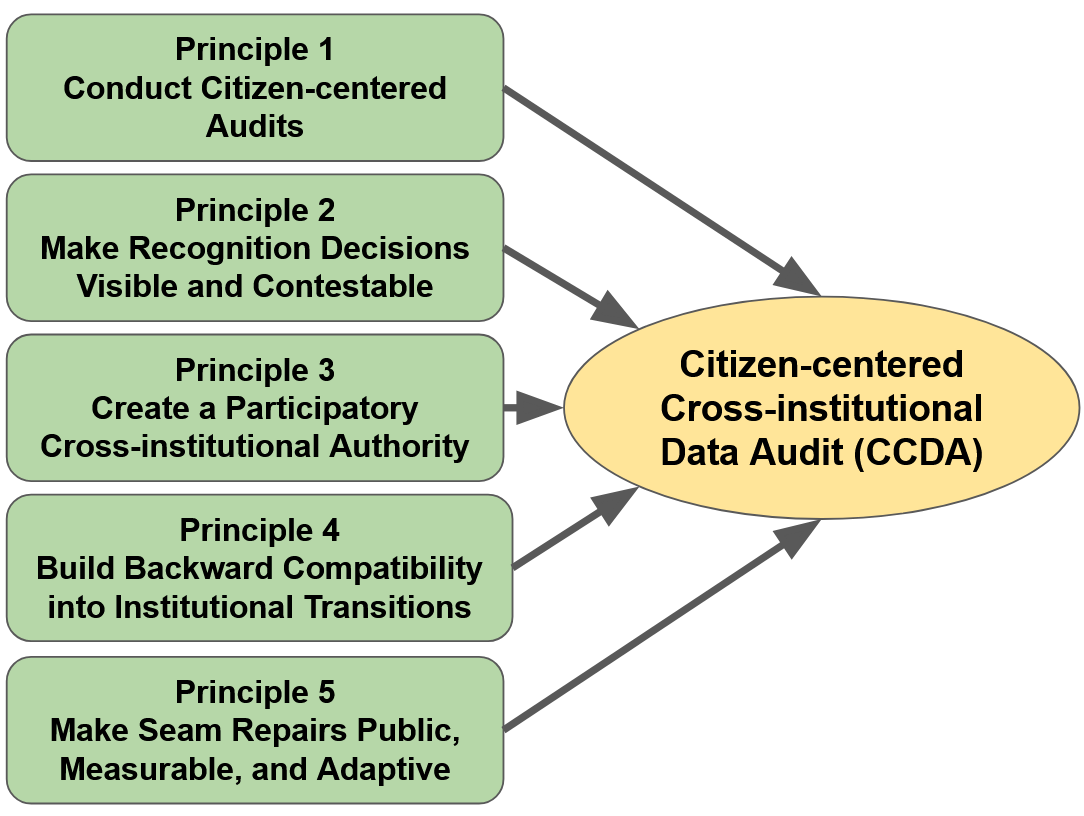}
    \vspace{-10pt}
    \caption{Five principles of the Citizen-centered Cross-institutional Data Audit (CCDA). 
    }
    \label{fig:five-principles}
    \vspace{-15pt}
\end{wrapfigure}

\subsubsection{Principle 1: Conduct Citizen-centered Audits}
Our first principle proposes a citizen-centered audit of institutional seams. The audit should start with citizens in a particular territory and trace the institutional crossings they experience. A seam audit should therefore compare three accounts: what the issuing institution records, what the receiving institution records, and what the person experienced while moving between them. This citizen-centered approach requires auditors to actively reach people rather than depend on institutional records or formal complaints. By directly documenting people's experiences, auditors can identify delays, additional visits, costs, repeated evidence, and informal payments that institutional records leave out \cite{watson2024hostile}. They can also compare these experiences across locations to identify geographic inequalities. For HCI, this means that evaluating a citizen service that involves multiple institutions should not stop at whether an individual portal, database, or institution works as designed. The audit should remain independent of the institutions being examined, and testimony should be protected when people report informal practices.

\subsubsection{Principle 2: Make Recognition Decisions Visible and Contestable}
Our second principle requires datafied citizen services to make recognition decisions visible. Our findings noted cases where institutions did not formally reject a record. They simply could not process it, requested more evidence, devalued it, or delayed the case. Without a recorded decision, these seam failures can remain invisible to both the citizen and future audits. We suggest that institutions should provide a documented reason in contested cases and state why the record was not accepted, what is needed for the process to continue, and how the citizen can contest the decision. Long delays and repeated requests for evidence should also be recorded. For HCI, this creates a design requirement for datafied citizen services. Interfaces and service systems should not only show whether a request is approved or rejected. Citizens should be able to see where their record stopped, why it stopped, what institution made the decision, and what they can do next.

\subsubsection{Principle 3: Create a Participatory Cross-institutional Authority}
Our third principle is that institutional seams should have an accountable, participatory authority with the power to coordinate and repair cross-institutional failures. Our findings (see 5.3) show that people could identify seam failures and attempt workarounds but lacked the authority to create shared verification standards or resolve responsibility across institutions. Governance by patching similarly distinguishes between identifying a failure and having the authority to fix it \cite{veeraraghavan2022patching}. CCDA therefore requires a ministry, regulator, inter-agency body, or other designated authority with jurisdiction over the seam; where none exists, that responsibility should be explicitly created. This authority should be able to establish equivalence between forms of evidence, negotiate shared standards, require verification or transfer pathways, assign responsibility for stranded records, and provide temporary routes while permanent solutions are developed. It should also involve affected citizens in decisions about seam repair. Auditing and repair should remain distinct: CCDA should document recurring failures, while the responsible authority acts on them. A shared register can connect these functions by recording failures, repairs, and accountable institutions or authorities.

\subsubsection{Principle 4: Build Backward Compatibility into Institutional Transitions}
Our fourth principle is that institutional transitions should preserve the usability of records created under earlier systems. Changes in technology, data formats, rules, administration, or jurisdiction should not make previously recognized records unusable. Backward compatibility should therefore be treated as both a technical and institutional requirement in datafied citizen services. When older records cannot be directly converted, institutions should maintain a working process for updating or revalidating them for as long as citizens still depend on those records. The same principle should apply when authority shifts between institutions or administrations: political or organizational transitions should not break verification of legitimately issued records. New databases, portals, standards, or administrative arrangements should be introduced with explicit continuity mechanisms that identify dependent legacy records and provide a route for keeping them actionable.

\subsubsection{Principle 5: Make Seam Repairs Public, Measurable, and Adaptive}
Our final principle is that seam repairs should be public, measurable, and adaptive rather than treated as complete once a new pathway is created. A repair may reproduce existing inequalities, such as a verification process available only in Dhaka, and should therefore be evaluated after implementation and revised when new problems emerge. The shared register proposed in Principle 3 should document what crossing was repaired, what evidence is accepted, who is responsible, and how citizens can challenge the arrangement. Most importantly, repair success should be measured by whether it reduces citizens' work, including repeated visits, delays, personal costs, duplicated evidence, and dependence on informal payments or intermediaries. CCDA should therefore revisit repaired seams to assess whether citizens' experiences improve. For HCI, this means evaluating whether changes to interfaces, workflows, or data exchange reduce cross-institutional burdens; for policy, authorities should remain accountable for the effects of repairs rather than treating a new rule or pathway as the endpoint.


Taken together, the CCDA framework advances HCI and critical studies of data infrastructures by providing design directions for making datafied citizen services more visible, contestable, connected, and responsive to citizens' experiences. For policy, it identifies the need for shared standards, clear jurisdiction, continuity requirements, and ongoing accountability across institutional boundaries. Together, these changes move responsibility for making institutions connect away from individual citizens and toward the institutions and authorities that govern the datafied citizen service ecosystem.

\subsection{Limitations and Future Work}
This study has several limitations. First, our analysis relies primarily on participants' accounts. We did not interview officials from all involved institutions or independently verify every rule, decision, or explanation, and institutional actors may describe some cases differently. Second, the study was conducted in Bangladesh with people who had experienced difficulties making records actionable across institutions. These experiences are shaped by Bangladesh's institutional and social context and may differ elsewhere. Recruitment through community networks, word of mouth, and snowball sampling may also have favored participants who could identify and describe these problems, and our targeted sampling does not indicate how common such failures are in the broader population. Third, covering six institutional domains allowed us to identify cross-domain patterns but limited in-depth analysis of any single system. Some accounts also concerned events from years earlier and may reflect incomplete recall. Fourth, our critique of individualistic institutional audits is conceptual and grounded in our findings and prior literature. We did not conduct audits or interview auditors, and the five CCDA principles have not yet been implemented or evaluated.

Future work should examine institutional seams from multiple perspectives by combining people's experiences with interviews with officials, institutional records, and observations of how records are accepted, rejected, transferred, and verified. Comparative studies across institutional and national contexts could identify which forms of cross-institutional data devaluation recur and which are context-specific. Longitudinal work could follow records across institutions in real time to examine particular seams in greater depth and reduce reliance on retrospective accounts. Finally, future research should implement and evaluate CCDA with institutions, auditors, policymakers, and affected communities to assess its practical challenges, unintended consequences, and effectiveness in reducing the work people must do to make institutions connect.

\section{Conclusion}
This paper introduces cross-institutional data devaluation to describe how valid records can lose recognition, credibility, or practical value when they move across institutional seams in datafied citizen services. Our interview study identifies three forms of devaluation: missing institutional connections, unequal recognition across existing connections, and connections that become obsolete after institutional change. We also show how people respond through citizen-borne workarounds, which may help an individual case move forward but does not repair the underlying institutional seam. Therefore, we propose the CCDA framework to make these seam failures visible, contestable, accountable, and repairable. Together, these contributions shift attention from whether individual institutions work correctly to whether the connections between them work for people who depend on citizen services that involve multiple institutions.




\bibliographystyle{ACM-Reference-Format}
\bibliography{sample-base}

@article{fereday2006demonstrating,
  title={Demonstrating rigor using thematic analysis: A hybrid approach of inductive and deductive coding and theme development},
  author={Fereday, Jennifer and Muir-Cochrane, Eimear},
  journal={International journal of qualitative methods},
  volume={5},
  number={1},
  pages={80--92},
  year={2006},
  publisher={SAGE Publications Sage CA: Los Angeles, CA}
}

@book{76boyatzis1998transforming,
  title={Transforming qualitative information: Thematic analysis and code development},
  author={Boyatzis, Richard E},
  year={1998},
  publisher={sage}
}

@book{Harper1998,
  author    = {Harper, Richard H. R.},
  title     = {Inside the {IMF}: An Ethnography of Documents, Technology and Organisational Action},
  publisher = {Academic Press},
  address   = {San Diego, CA},
  year      = {1998}
}

@article{Hull2012,
  author  = {Hull, Matthew S.},
  title   = {Documents and Bureaucracy},
  journal = {Annual Review of Anthropology},
  volume  = {41},
  number  = {1},
  pages   = {251--267},
  year    = {2012},
  doi     = {10.1146/annurev.anthro.012809.104953}
}

@book{SharmaGupta2006,
  editor    = {Sharma, Aradhana and Gupta, Akhil},
  title     = {The Anthropology of the State: A Reader},
  publisher = {Blackwell Publishing},
  address   = {Malden, MA},
  year      = {2006}
}

@book{Porter1995,
  author    = {Porter, Theodore M.},
  title     = {Trust in Numbers: The Pursuit of Objectivity in Science and Public Life},
  publisher = {Princeton University Press},
  address   = {Princeton, NJ},
  year      = {1995}
}

@book{Scott1998,
  author    = {Scott, James C.},
  title     = {Seeing Like a State: How Certain Schemes to Improve the Human Condition Have Failed},
  publisher = {Yale University Press},
  address   = {New Haven, CT},
  year      = {1998}
}

@book{BowkerStar1999,
  author    = {Bowker, Geoffrey C. and Star, Susan Leigh},
  title     = {Sorting Things Out: Classification and Its Consequences},
  publisher = {MIT Press},
  address   = {Cambridge, MA},
  year      = {1999}
}

@article{StarBowker2007,
  author  = {Star, Susan Leigh and Bowker, Geoffrey C.},
  title   = {Enacting Silence: Residual Categories as a Challenge for Ethics, Information Systems, and Communication},
  journal = {Ethics and Information Technology},
  volume  = {9},
  number  = {4},
  pages   = {273--280},
  year    = {2007},
  doi     = {10.1007/s10676-007-9141-7}
}

@article{MartinLynch2009,
  author  = {Martin, Aryn and Lynch, Michael},
  title   = {Counting Things and People: The Practices and Politics of Counting},
  journal = {Social Problems},
  volume  = {56},
  number  = {2},
  pages   = {243--266},
  year    = {2009},
  doi     = {10.1525/sp.2009.56.2.243}
}

@book{BreckenridgeSzreter2012,
  editor    = {Breckenridge, Keith and Szreter, Simon},
  title     = {Registration and Recognition: Documenting the Person in World History},
  publisher = {Oxford University Press for the British Academy},
  address   = {Oxford, UK},
  year      = {2012}
}

@book{CaplanTorpey2001,
  editor    = {Caplan, Jane and Torpey, John},
  title     = {Documenting Individual Identity: The Development of State Practices in the Modern World},
  publisher = {Princeton University Press},
  address   = {Princeton, NJ},
  year      = {2001}
}

@article{Sriraman2011,
  author  = {Sriraman, Tarangini},
  title   = {Revisiting Welfare: Ration Card Narratives in India},
  journal = {Economic and Political Weekly},
  volume  = {46},
  number  = {38},
  pages   = {52--59},
  year    = {2011}
}

@inproceedings{SinghJackson2017,
  author    = {Singh, Ranjit and Jackson, Steven J.},
  title     = {From Margins to Seams: Imbrication, Inclusion, and Torque in the {Aadhaar} Identification Project},
  booktitle = {Proceedings of the 2017 CHI Conference on Human Factors in Computing Systems},
  series    = {CHI '17},
  pages     = {4776--4824},
  year      = {2017},
  publisher = {Association for Computing Machinery},
  address   = {New York, NY, USA},
  doi       = {10.1145/3025453.3025910}
}

@article{SinghJackson2021,
  author  = {Singh, Ranjit and Jackson, Steven J.},
  title   = {Seeing Like an Infrastructure: Low-Resolution Citizens and the {Aadhaar} Identification Project},
  journal = {Proceedings of the ACM on Human-Computer Interaction},
  volume  = {5},
  number  = {CSCW2},
  articleno = {315},
  numpages = {26},
  pages   = {1--26},
  year    = {2021},
  publisher = {Association for Computing Machinery},
  address = {New York, NY, USA},
  doi     = {10.1145/3476056}
}

@article{EspelandStevens1998,
  author  = {Espeland, Wendy Nelson and Stevens, Mitchell L.},
  title   = {Commensuration as a Social Process},
  journal = {Annual Review of Sociology},
  volume  = {24},
  number  = {1},
  pages   = {313--343},
  year    = {1998},
  doi     = {10.1146/annurev.soc.24.1.313}
}

@article{Lamont2012,
  author  = {Lamont, Mich{\`e}le},
  title   = {Toward a Comparative Sociology of Valuation and Evaluation},
  journal = {Annual Review of Sociology},
  volume  = {38},
  number  = {1},
  pages   = {201--221},
  year    = {2012},
  doi     = {10.1146/annurev-soc-070308-120022}
}

@article{FourcadeHealy2013,
  author  = {Fourcade, Marion and Healy, Kieran},
  title   = {Classification Situations: Life-Chances in the Neoliberal Era},
  journal = {Accounting, Organizations and Society},
  volume  = {38},
  number  = {8},
  pages   = {559--572},
  year    = {2013},
  doi     = {10.1016/j.aos.2013.11.002}
}

@article{Fourcade2021,
  author  = {Fourcade, Marion},
  title   = {Ordinal Citizenship},
  journal = {The British Journal of Sociology},
  volume  = {72},
  number  = {2},
  pages   = {154--173},
  year    = {2021},
  doi     = {10.1111/1468-4446.12839}
}

@inproceedings{RohanifarEtAl2026,
  author    = {Rohanifar, Yasaman and Levine, Rachel F. J. and Sultana, Sharifa and Ahmed, Syed Ishtiaque},
  title     = {Data Migration in {HCI}: The Politics of Invisible Borders, Informal Networks, and Immigrants' Data},
  booktitle = {Proceedings of the 2026 CHI Conference on Human Factors in Computing Systems},
  series    = {CHI '26},
  year      = {2026},
  publisher = {Association for Computing Machinery},
  address   = {New York, NY, USA},
  doi       = {10.1145/3772318.3790605}
}

@article{StarRuhleder1996,
  author  = {Star, Susan Leigh and Ruhleder, Karen},
  title   = {Steps Toward an Ecology of Infrastructure: Design and Access for Large Information Spaces},
  journal = {Information Systems Research},
  volume  = {7},
  number  = {1},
  pages   = {111--134},
  year    = {1996},
  doi     = {10.1287/isre.7.1.111}
}

@article{Star1999,
  author  = {Star, Susan Leigh},
  title   = {The Ethnography of Infrastructure},
  journal = {American Behavioral Scientist},
  volume  = {43},
  number  = {3},
  pages   = {377--391},
  year    = {1999},
  doi     = {10.1177/00027649921955326}
}

@article{HansethLundberg2001,
  author  = {Hanseth, Ole and Lundberg, Nina},
  title   = {Designing Work Oriented Infrastructures},
  journal = {Computer Supported Cooperative Work (CSCW)},
  volume  = {10},
  number  = {3--4},
  pages   = {347--372},
  year    = {2001},
  doi     = {10.1023/A:1012727708439}
}

@article{PipekWulf2009,
  author  = {Pipek, Volkmar and Wulf, Volker},
  title   = {Infrastructuring: Toward an Integrated Perspective on the Design and Use of Information Technology},
  journal = {Journal of the Association for Information Systems},
  volume  = {10},
  number  = {5},
  pages   = {447--473},
  year    = {2009},
  doi     = {10.17705/1jais.00195}
}

@book{LamplandStar2009,
  editor    = {Lampland, Martha and Star, Susan Leigh},
  title     = {Standards and Their Stories: How Quantifying, Classifying, and Formalizing Practices Shape Everyday Life},
  publisher = {Cornell University Press},
  address   = {Ithaca, NY},
  year      = {2009}
}

@article{StarGriesemer1989,
  author  = {Star, Susan Leigh and Griesemer, James R.},
  title   = {Institutional Ecology, `Translations' and Boundary Objects: Amateurs and Professionals in Berkeley's Museum of Vertebrate Zoology, 1907--39},
  journal = {Social Studies of Science},
  volume  = {19},
  number  = {3},
  pages   = {387--420},
  year    = {1989},
  doi     = {10.1177/030631289019003001}
}

@article{TimmermansEpstein2010,
  author  = {Timmermans, Stefan and Epstein, Steven},
  title   = {A World of Standards but not a Standard World: Toward a Sociology of Standards and Standardization},
  journal = {Annual Review of Sociology},
  volume  = {36},
  number  = {1},
  pages   = {69--89},
  year    = {2010},
  doi     = {10.1146/annurev.soc.012809.102629}
}

@inproceedings{JacksonChongthammakun2011,
  author    = {Jackson, Steven J. and Chongthammakun, Radaphat},
  title     = {Infrastructure and Standards in {Thai} Digital Government},
  booktitle = {Proceedings of the 2011 iConference},
  series    = {iConference '11},
  pages     = {379--386},
  year      = {2011},
  publisher = {Association for Computing Machinery},
  address   = {New York, NY, USA},
  doi       = {10.1145/1940761.1940813}
}

@article{Vertesi2014,
  author  = {Vertesi, Janet},
  title   = {Seamful Spaces: Heterogeneous Infrastructures in Interaction},
  journal = {Science, Technology, \& Human Values},
  volume  = {39},
  number  = {2},
  pages   = {264--284},
  year    = {2014},
  doi     = {10.1177/0162243913516012}
}

@article{SchmidtSimone1996,
  author  = {Schmidt, Kjeld and Simone, Carla},
  title   = {Coordination Mechanisms: Towards a Conceptual Foundation of {CSCW} Systems Design},
  journal = {Computer Supported Cooperative Work (CSCW)},
  volume  = {5},
  number  = {2--3},
  pages   = {155--200},
  year    = {1996},
  doi     = {10.1007/BF00133655}
}

@article{StarStrauss1999,
  author  = {Star, Susan Leigh and Strauss, Anselm},
  title   = {Layers of Silence, Arenas of Voice: The Ecology of Visible and Invisible Work},
  journal = {Computer Supported Cooperative Work (CSCW)},
  volume  = {8},
  number  = {1--2},
  pages   = {9--30},
  year    = {1999},
  doi     = {10.1023/A:1008651105359}
}

@inproceedings{AhmedEtAl2015,
  author    = {Ahmed, Syed Ishtiaque and Mim, Nusrat Jahan and Jackson, Steven J.},
  title     = {Residual Mobilities: Infrastructural Displacement and Post-Colonial Computing in {Bangladesh}},
  booktitle = {Proceedings of the 33rd Annual ACM Conference on Human Factors in Computing Systems},
  series    = {CHI '15},
  pages     = {437--446},
  year      = {2015},
  publisher = {Association for Computing Machinery},
  address   = {New York, NY, USA},
  doi       = {10.1145/2702123.2702573}
}

@article{RangaswamySambasivan2011,
  author  = {Rangaswamy, Nimmi and Sambasivan, Nimmi},
  title   = {Cutting Chai, Jugaad, and Here Pheri: Towards {UbiComp} for a Global Community},
  journal = {Personal and Ubiquitous Computing},
  volume  = {15},
  number  = {6},
  pages   = {553--564},
  year    = {2011},
  doi     = {10.1007/s00779-010-0349-x}
}

@article{Chaudhuri2019,
  author  = {Chaudhuri, Bidisha},
  title   = {Paradoxes of Intermediation in {Aadhaar}: Human Making of a Digital Infrastructure},
  journal = {South Asia: Journal of South Asian Studies},
  volume  = {42},
  number  = {3},
  pages   = {572--587},
  year    = {2019},
  doi     = {10.1080/00856401.2019.1598671}
}

@article{CarswellDeNeve2021,
  author  = {Carswell, Grace and De Neve, Geert},
  title   = {Transparency, Exclusion and Mediation: How Digital and Biometric Technologies are Transforming Social Protection in {Tamil Nadu}, {India}},
  journal = {Oxford Development Studies},
  volume  = {50},
  number  = {2},
  pages   = {126--141},
  year    = {2022},
  doi     = {10.1080/13600818.2021.1904866}
}

@incollection{Jackson2014,
  author    = {Jackson, Steven J.},
  title     = {Rethinking Repair},
  booktitle = {Media Technologies: Essays on Communication, Materiality, and Society},
  editor    = {Gillespie, Tarleton and Boczkowski, Pablo J. and Foot, Kirsten A.},
  pages     = {221--239},
  publisher = {MIT Press},
  address   = {Cambridge, MA},
  year      = {2014},
  doi       = {10.7551/mitpress/9780262525374.003.0011}
}

@book{MarquandHouseCollective2026,
  author    = {{The Marquand House Collective}},
  title     = {Auditing {AI}},
  series    = {MIT Press Essential Knowledge},
  publisher = {MIT Press},
  address   = {Cambridge, MA},
  year      = {2026},
  isbn      = {9780262051729}
}

@article{MetaxaEtAl2021AuditingAlgorithms,
  author  = {Metaxa, Dana{\"e} and Park, Joon Sung and Robertson, Ronald E. and Karahalios, Karrie and Wilson, Christo and Hancock, Jeff and Sandvig, Christian},
  title   = {Auditing Algorithms: Understanding Algorithmic Systems from the Outside In},
  journal = {Foundations and Trends in Human-Computer Interaction},
  volume  = {14},
  number  = {4},
  pages   = {272--344},
  year    = {2021},
  doi     = {10.1561/1100000083}
}

@inproceedings{SandvigEtAl2014,
  author    = {Sandvig, Christian and Hamilton, Kevin and Karahalios, Karrie and Langbort, Cedric},
  title     = {Auditing Algorithms: Research Methods for Detecting Discrimination on Internet Platforms},
  booktitle = {Data and Discrimination: Converting Critical Concerns into Productive Inquiry, a Preconference at the 64th Annual Meeting of the International Communication Association},
  year      = {2014},
  address   = {Seattle, WA, USA}
}

@article{PagerWestern2012,
  author  = {Pager, Devah and Western, Bruce},
  title   = {Identifying Discrimination at Work: The Use of Field Experiments},
  journal = {Journal of Social Issues},
  volume  = {68},
  number  = {2},
  pages   = {221--237},
  year    = {2012},
  doi     = {10.1111/j.1540-4560.2012.01746.x}
}

@book{Gaddis2018,
  editor    = {Gaddis, S. Michael},
  title     = {Audit Studies: Behind the Scenes with Theory, Method, and Nuance},
  publisher = {Springer},
  address   = {Cham, Switzerland},
  year      = {2018},
  doi       = {10.1007/978-3-319-71153-9}
}

@article{BertrandMullainathan2004,
  author  = {Bertrand, Marianne and Mullainathan, Sendhil},
  title   = {Are {Emily} and {Greg} More Employable Than {Lakisha} and {Jamal}? A Field Experiment on Labor Market Discrimination},
  journal = {American Economic Review},
  volume  = {94},
  number  = {4},
  pages   = {991--1013},
  year    = {2004},
  doi     = {10.1257/0002828042002561}
}

@inproceedings{AsplundEtAl2020,
  author    = {Asplund, Jakob and Eslami, Motahhare and Sundaram, Hari and Sandvig, Christian and Karahalios, Karrie},
  title     = {Auditing Race and Gender Discrimination in Online Housing Markets},
  booktitle = {Proceedings of the International AAAI Conference on Web and Social Media},
  volume    = {14},
  pages     = {24--35},
  year      = {2020},
  doi       = {10.1609/icwsm.v14i1.7276}
}

@article{Sweeney2013,
  author  = {Sweeney, Latanya},
  title   = {Discrimination in Online Ad Delivery},
  journal = {Communications of the ACM},
  volume  = {56},
  number  = {5},
  pages   = {44--54},
  year    = {2013},
  doi     = {10.1145/2447976.2447990}
}

@inproceedings{HannakEtAl2014,
  author    = {Hannak, Aniko and Soeller, Gary and Lazer, David and Mislove, Alan and Wilson, Christo},
  title     = {Measuring Price Discrimination and Steering on E-commerce Web Sites},
  booktitle = {Proceedings of the 2014 Conference on Internet Measurement Conference},
  series    = {IMC '14},
  pages     = {305--318},
  year      = {2014},
  publisher = {Association for Computing Machinery},
  address   = {New York, NY, USA},
  doi       = {10.1145/2663716.2663744}
}

@inproceedings{BuolamwiniGebru2018,
  author    = {Buolamwini, Joy and Gebru, Timnit},
  title     = {Gender Shades: Intersectional Accuracy Disparities in Commercial Gender Classification},
  booktitle = {Proceedings of the 1st Conference on Fairness, Accountability and Transparency},
  series    = {Proceedings of Machine Learning Research},
  volume    = {81},
  pages     = {77--91},
  year      = {2018},
  publisher = {PMLR}
}

@inproceedings{RajiBuolamwini2019,
  author    = {Raji, Inioluwa Deborah and Buolamwini, Joy},
  title     = {Actionable Auditing: Investigating the Impact of Publicly Naming Biased Performance Results of Commercial {AI} Products},
  booktitle = {Proceedings of the 2019 AAAI/ACM Conference on AI, Ethics, and Society},
  series    = {AIES '19},
  pages     = {429--435},
  year      = {2019},
  publisher = {Association for Computing Machinery},
  address   = {New York, NY, USA},
  doi       = {10.1145/3306618.3314244}
}

@inproceedings{KayEtAl2015,
  author    = {Kay, Matthew and Matuszek, Cynthia and Munson, Sean A.},
  title     = {Unequal Representation and Gender Stereotypes in Image Search Results for Occupations},
  booktitle = {Proceedings of the 33rd Annual ACM Conference on Human Factors in Computing Systems},
  series    = {CHI '15},
  pages     = {3819--3828},
  year      = {2015},
  publisher = {Association for Computing Machinery},
  address   = {New York, NY, USA},
  doi       = {10.1145/2702123.2702520}
}

@article{MetaxaEtAl2021ImageSociety,
  author  = {Metaxa, Dana{\"e} and Gan, Michelle A. and Goh, Su and Hancock, Jeff and Landay, James A.},
  title   = {An Image of Society: Gender and Racial Representation and Impact in Image Search Results for Occupations},
  journal = {Proceedings of the ACM on Human-Computer Interaction},
  volume  = {5},
  number  = {CSCW1},
  articleno = {26},
  year    = {2021},
  publisher = {Association for Computing Machinery},
  address = {New York, NY, USA},
  doi     = {10.1145/3449100}
}

@misc{AngwinParris2016,
  author       = {Angwin, Julia and Parris, Jr., Terry},
  title        = {Facebook Lets Advertisers Exclude Users by Race},
  howpublished = {ProPublica},
  month        = oct,
  year         = {2016},
  note         = {\url{https://www.propublica.org/article/facebook-lets-advertisers-exclude-users-by-race}}
}

@misc{AngwinTobinVarner2017,
  author       = {Angwin, Julia and Tobin, Ariana and Varner, Madeleine},
  title        = {Facebook (Still) Letting Housing Advertisers Exclude Users by Race},
  howpublished = {ProPublica},
  month        = nov,
  year         = {2017},
  note         = {\url{https://www.propublica.org/article/facebook-advertising-discrimination-housing-race-sex-national-origin}}
}

@article{AliEtAl2019,
  author  = {Ali, Muhammad and Sapiezynski, Piotr and Bogen, Miranda and Korolova, Aleksandra and Mislove, Alan and Rieke, Aaron},
  title   = {Discrimination Through Optimization: How {Facebook's} Ad Delivery Can Lead to Biased Outcomes},
  journal = {Proceedings of the ACM on Human-Computer Interaction},
  volume  = {3},
  number  = {CSCW},
  articleno = {199},
  year    = {2019},
  publisher = {Association for Computing Machinery},
  address = {New York, NY, USA},
  doi     = {10.1145/3359301}
}

@inproceedings{ArmstrongEtAl2024,
  author    = {Armstrong, Lena and Liu, Abbey and MacNeil, Stephen and Metaxa, Dana{\"e}},
  title     = {The Silicon Ceiling: Auditing {GPT's} Race and Gender Biases in Hiring},
  booktitle = {Proceedings of the 4th ACM Conference on Equity and Access in Algorithms, Mechanisms, and Optimization},
  series    = {EAAMO '24},
  year      = {2024},
  publisher = {Association for Computing Machinery},
  address   = {New York, NY, USA},
  doi       = {10.1145/3689904.3694699}
}

@article{ShenEtAl2021,
  author  = {Shen, Hong and DeVos, Alicia and Eslami, Motahhare and Holstein, Kenneth},
  title   = {Everyday Algorithm Auditing: Understanding the Power of Everyday Users in Surfacing Harmful Algorithmic Behaviors},
  journal = {Proceedings of the ACM on Human-Computer Interaction},
  volume  = {5},
  number  = {CSCW2},
  articleno = {433},
  year    = {2021},
  publisher = {Association for Computing Machinery},
  address = {New York, NY, USA},
  doi     = {10.1145/3479577}
}

@article{LamEtAl2022,
  author  = {Lam, Michelle S. and Gordon, Mitchell L. and Metaxa, Dana{\"e} and Hancock, Jeffrey T. and Landay, James A. and Bernstein, Michael S.},
  title   = {End-User Audits: A System Empowering Communities to Lead Large-Scale Investigations of Harmful Algorithmic Behavior},
  journal = {Proceedings of the ACM on Human-Computer Interaction},
  volume  = {6},
  number  = {CSCW2},
  articleno = {512},
  year    = {2022},
  publisher = {Association for Computing Machinery},
  address = {New York, NY, USA},
  doi     = {10.1145/3555625}
}

@inproceedings{GrovesEtAl2024,
  author    = {Groves, Lara and Metcalf, Jacob and Kennedy, Alayna and Vecchione, Briana and Strait, Andrew},
  title     = {Auditing Work: Exploring the New York City Algorithmic Bias Audit Regime},
  booktitle = {Proceedings of the 2024 ACM Conference on Fairness, Accountability, and Transparency},
  series    = {FAccT '24},
  year      = {2024},
  publisher = {Association for Computing Machinery},
  address   = {New York, NY, USA},
  doi       = {10.1145/3630106.3658959}
}

@inproceedings{WrightEtAl2024,
  author    = {Wright, Lucas and Muenster, Roxana Mika and Vecchione, Briana and Qu, Tianyao and Cai, Senhuang and Smith, Alan and Metcalf, Jacob and Matias, J. Nathan},
  title     = {Null Compliance: {NYC} Local Law 144 and the Challenges of Algorithm Accountability},
  booktitle = {Proceedings of the 2024 ACM Conference on Fairness, Accountability, and Transparency},
  series    = {FAccT '24},
  year      = {2024},
  publisher = {Association for Computing Machinery},
  address   = {New York, NY, USA},
  doi       = {10.1145/3630106.3658972}
}

@inproceedings{WilsonEtAl2021,
  author    = {Wilson, Christo and Ghosh, Avijit and Jiang, Shan and Mislove, Alan and Baker, Lewis and Szary, Janelle and Trindel, Kelly and Polli, Frida},
  title     = {Building and Auditing Fair Algorithms: A Case Study in Candidate Screening},
  booktitle = {Proceedings of the 2021 ACM Conference on Fairness, Accountability, and Transparency},
  series    = {FAccT '21},
  pages     = {666--677},
  year      = {2021},
  publisher = {Association for Computing Machinery},
  address   = {New York, NY, USA},
  doi       = {10.1145/3442188.3445928}
}

@inproceedings{YoungEtAl2022,
  author    = {Young, Meg and Katell, Michael and Krafft, P. M.},
  title     = {Confronting Power and Corporate Capture at the {FAccT} Conference},
  booktitle = {Proceedings of the 2022 ACM Conference on Fairness, Accountability, and Transparency},
  series    = {FAccT '22},
  pages     = {1375--1386},
  year      = {2022},
  publisher = {Association for Computing Machinery},
  address   = {New York, NY, USA},
  doi       = {10.1145/3531146.3533194}
}

@book{veeraraghavan2022patching,
  title={Patching development: Information politics and social change in India},
  author={Veeraraghavan, Rajesh},
  year={2022},
  publisher={Oxford University Press Oxford}
}

@inproceedings{selbst2019fairness,
  title={Fairness and abstraction in sociotechnical systems},
  author={Selbst, Andrew D and Boyd, Danah and Friedler, Sorelle A and Venkatasubramanian, Suresh and Vertesi, Janet},
  booktitle={Proceedings of the conference on fairness, accountability, and transparency},
  pages={59--68},
  year={2019}
}

@inproceedings{cooper2022accountability,
  title={Accountability in an algorithmic society: relationality, responsibility, and robustness in machine learning},
  author={Cooper, A Feder and Moss, Emanuel and Laufer, Benjamin and Nissenbaum, Helen},
  booktitle={Proceedings of the 2022 ACM conference on fairness, accountability, and transparency},
  pages={864--876},
  year={2022}
}

@inproceedings{SultanaAhmed2026CrossInstitutional,
  author    = {Sharifa Sultana and Syed Ishtiaque Ahmed},
  title     = {Cross-Institutional AI Audits},
  booktitle = {Proceedings of the ACM AI Summit 2026 (AI Summit '26)},
  year      = {2026},
  address   = {Atlanta, GA, USA},
  publisher = {Association for Computing Machinery},
  doi       = {10.1145/3806096.3844800}
}

@inproceedings{coles2019accessing,
  title={Accessing a new land: Designing for a social conceptualisation of access},
  author={Coles-Kemp, Lizzie and Jensen, Rikke Bjerg},
  booktitle={Proceedings of the 2019 CHI Conference on Human Factors in Computing Systems},
  pages={1--12},
  year={2019}
}

@inproceedings{rohanifar2021money,
  title={Money whispers: Informality, international politics, and immigration in transnational finance},
  author={Rohanifar, Yasaman and Chandra, Priyank and Rahman, M Ataur and Ahmed, Syed Ishtiaque},
  booktitle={Proceedings of the 2021 CHI conference on human factors in computing systems},
  pages={1--14},
  year={2021}
}

@inproceedings{watson2024hostile,
  title={Hostile Systems: A Taxonomy of Harms Articulated by Citizens Living with Socio-Economic Deprivation},
  author={Watson, Colin and Crivellaro, Clara and Parnaby, Adam W and Kharrufa, Ahmed},
  booktitle={Proceedings of the 2024 CHI Conference on Human Factors in Computing Systems},
  pages={1--17},
  year={2024}
}

\end{document}